# Nucleation of stoichiometric compounds from liquid: the role of the kinetic factor

H. Song[a], Y. Sun[a], F, Zhang[a], C.Z. Wang[a], K.M. Ho[a,b,c] and M.I. Mendelev[a]

[a] Division of Materials Sciences and Engineering, Ames Laboratory (US Department of Energy), Ames, Iowa 50011, USA
[b] Department of Physics, Iowa State University, Ames, Iowa 50011, USA
[c] Hefei National Laboratory for Physical Sciences at the Microscale and Department of Physics, University of Science and Technology of China, Hefei, Anhui 230026, China

**Abstract**

While the role of the free energy barrier during nucleation is a text-book subject the importance of the kinetic factor is frequently underestimated. We obtained both quantities from molecular dynamics (MD) simulations for the pure Ni and B2 phases in the $Ni_{50}Al_{50}$ and $Cu_{50}Zr_{50}$ alloys. The free-energy barrier was found to be higher in Ni but the nucleation rate is much lower in the $Ni_{50}Al_{50}$ alloy which was attributed to the ordered nature of the B2 phase. Since the $Cu_{50}Zr_{50}$ B2 phase can has even smaller fraction of the anti-site defects its nucleation is never observed in the MD simulation.

Crystal nucleation from the liquid has been extensively studied in the several past decades [1, 2]. According to the classical nucleation theory (CNT) [3], forming a crystalline nucleus can be described as a competition between the bulk driving force and the energy penalty associated with creating an interface between the nucleus and liquid. The excess free energy to form a nucleus with $n$ atoms can be expressed as $\Delta G = n\Delta\mu + A\gamma$, where $\Delta\mu$ ($< 0$) is the difference between the bulk solid and liquid free energies, $\gamma$ is the solid-liquid interfacial (SLI) free energy, and $A$ is the interface area. When the crystal nucleus is small its growth leads to increasing the free energy but once the nucleus is larger than the critical size, $n^*$, its growth leads to decreasing the free energy. Thus, the excess of the free energy, $\Delta G^*$, necessary to form the nucleus of the critical size is considered as the nucleation barrier. The nucleation rate can be written as

$$J = \kappa \cdot exp\left(-\frac{\Delta G^*}{k_B T}\right) \quad (1)$$

where

$$\kappa = \rho_L f_{n^*}^+ Z \quad (2)$$

is the kinetic prefactor, which depends on the atomic density of the liquid phase $\rho_L$, the rate of attachment of individual atoms to the critical cluster $f_{n^*}^+$, and the Zeldovich factor $Z$ [4], which describes the curvature of free energy landscape at the top of the barrier $Z = \sqrt{|\Delta G''(n^*)|/2\pi k_B T}$.

It is the free energy barrier which is usually considered to dominate the nucleation rate $J$. Therefore, it is frequently used to predict whether the nucleation will proceed fast or not. If we assume that the nucleus has a spherical shape, the nucleation energy barrier, $\Delta G^*$, can be calculated as

$$\Delta G_{CNT}^{S*} = \frac{16}{3}\pi \frac{\gamma^3}{\Delta \mu^2}. \tag{3}$$

Modern simulation techniques allow to obtain reliable data on the bulk driving forces, $\Delta \mu$ [5-7] and the SLI free energies at the melting temperatures, $\gamma$ [8, 9]. *However, how realistic are the predictions made based just on these data?* Consider as example the crystal nucleation in three systems described by the embedded atom method [10] (EAM) and the Finnis-Sinclair [11] potentials: pure Ni [12] (which has face-centered cubic (fcc) lattice) and B2 phases in the $Ni_{50}Al_{50}$ [13] and $Cu_{50}Zr_{50}$ [11-14] alloys. The bulk driving force and the SLI free energy were obtained for all of these systems from MD simulations [15, 16] and $\Delta G_{CNT}^{S*}$ is shown in Fig. 1. Examination of these data suggests that the nucleation will be the slowest in the pure Ni; it will proceed a little faster in the $Cu_{50}Zr_{50}$ alloy and very fast in the $Ni_{50}Al_{50}$ alloy. The MD simulation results are in vivid contradiction with this prediction: while no nucleation has been observed in the $Cu_{50}Zr_{50}$ alloy [17], it is readily observed in the pure Ni as we will show below. We note that these MD simulation results are consistent with experimental observations: the $Cu_{50}Zr_{50}$ alloy is a good glass former [18] while no amorphous Ni sample has been synthesized. Moreover, the MD simulation shows that the nucleation in the pure Ni proceeds much faster than in the $Ni_{50}Al_{50}$ alloy (see below) while the data shown in Fig. 1 suggest the opposite. Thus, even using very accurate input data does not allow to make any realistic predictions based on Eq. (3).

Equation (3) is derived using several rather strong approximations: the nucleus is assumed to be spherical and the SLI free energy is assumed to be temperature and curvature independent. Recently, a considerable progress was achieved to evaluate activation barriers from simulation without making such approximations. For example, umbrella sampling [19, 20] or transition path sampling [21] methods map out the equilibrium free energy landscape for nucleation by adding a biased potential to the interatomic interaction. Recently, Wedekind et al. [22-24] developed the mean first-passage time (MFPT) method useful in the regime where nucleation can be observed in an unbiased MD simulation. The MFPT method yields the nucleation rate, the size of the critical embryo and the Zeldovich factor. Additionally, the Fokker-Planck (FP) equation [23, 25] can be used to analyze the same MD simulation data and reconstruct the free energy landscape as well as the attachment rates as a function of embryo size.

In this Letter, we report the results of the systematic MD simulation study of the homogeneous nucleation in the pure Ni and $Ni_{50}Al_{50}$ and $Cu_{50}Zr_{50}$ alloys using the EAM/FS potentials developed in [12, 26] and [14], respectively. The liquid models (32,000 atoms for the pure Ni and 54,000 atoms for the alloys) equilibrated at $T_m$ during 2 nanoseconds (ns) were taken as the initial configurations ($t_0$), and the simulations were started by quenching the liquid to the chosen undercooling temperature ($\Delta T$=540 K) which was approximately 30-40% of $T_m$. The 200 independent simulations were performed for each system by setting the different initial velocity random at configuration $t_0$. Finally, additional simulation cells containing up to 256,000 atoms were used to examine the system size effect on the simulation results. To define crystal-like embryos, we employed the widely-used bond-orientational order parameter (BOO) [19, 27, 28], with the threshold carefully chosen based on Espinosa *et al.*'s "equal mislabeling" method [29] for each simulation system. The crystal embryos were accurately defined by the ten Wolde-Frenkel definition [27, 28].

The MFPT method requires determination of the time when the *largest* crystal cluster in the system reaches or exceeds size $n$ for the first time (first-passage time, $\tau$). In the present study, we output snapshots every 50 fs and determined the number of atoms in the largest cluster. The simulation was terminated once the largest clusters exceeded 1000 atoms. It took from 100 ps to 150 ns to reach this state in the cases of the pure Ni and the $Ni_{50}Al_{50}$ alloy. However, for the $Cu_{50}Zr_{50}$ alloy, although small sub-clusters (i.e. crystal clusters with $n < n^*$) were detected in the simulations, we never observed any nucleation during 200 ns which is in clear contradiction with the predictions made based on Fig. 1. Thus, the MFPT method could not be applied to this alloy.

Examples of nucleation are shown in the insets of Fig. 2. In all simulation runs, we only observed one cluster passed the critical size at the moment. Typically, the pure Ni model contained more sub-clusters than did the $Ni_{50}Al_{50}$ alloy model. The critical nuclei in pure Ni were usually almost spherical while the critical nuclei in the $Ni_{50}Al_{50}$ alloy were very anisotropic. The time required to completely solidify the $Ni_{50}Al_{50}$ alloy model was much longer than that for the pure Ni.

Figure 2 shows functions $\tau(n)$ obtained by averaging over 200 crystallization runs. Due to sufficiently high nucleation barriers, these functions have a characteristic sigmoidal shape [22] with a clear plateau defined by the nucleation time $\tau_J$. The MFPT can be expressed as function of the largest cluster size as

$$\tau(n) = \frac{\tau_J}{2}\left[1 + erf\left(Z\sqrt{\pi}(n - n^*)\right)\right]. \qquad (4)$$

Fitting the simulation data shown in Fig. 2 to this equation allowed us to determine the nucleation time, $\tau_J$, the critical cluster size $n^*$ and the Zeldovich factor, $Z$. The obtained values are presented in Fig. 2 and Table I. The examination of this table reveals that the critical nucleus size in the pure Ni is almost twice larger than that in the $Ni_{50}Al_{50}$ alloy which is consistent with the prediction made based on Fig. 1. However, the nucleation rate in the pure Ni is 10 times larger than that in the $Ni_{50}Al_{50}$ alloy.

To explain this result, we need to separate the contributions of the nucleation barrier and the atom attachment kinetics to the nucleation rate. Wedeking and Reguera [23, 30] developed a method to reconstruct the free energy landscape and determine the attachment rate by using the MFPT data and the steady-state probability distribution of the largest nucleus in the system. Both of these ingredients can be directly obtained from MD simulation of nucleation. We note that the probability distribution of the largest nucleus $P_{larg}^{st}(n)$ is *unequal* to the full probability distribution of all nuclei, $P^{st}(n)$ in the system [31], but if we assume that the nucleation barrier is sufficiently high, these quantities can be related as: $P_{larg}^{st}(n) \simeq P^{st}(n) \cdot N$, where $N$ is the total number of atoms in the system. Then the free energy profile for formation of an individual cluster $\Delta G(n)$ and the free energy of the largest cluster $\Delta G_{larg}(n)$ can be related as $\Delta G(n) = \Delta G_{larg}(n) + k_B T \ln(N)$ (for more details see [23, 30, 32] and Supplementary Material). Figure 3(a) shows the reconstructed true free energy curve $\Delta G(n)/k_B T$ as a function of size $n$. This analysis leads to $\Delta G^*/k_B T$=19.5 and $n^*$ = 53 for the pure Ni, and $\Delta G^*/k_B T$=17.9 and $n^*$ = 28 for $Ni_{50}Al_{50}$. The obtained values of the critical nucleus size are different by 2 ~ 4 atoms comparing to the corresponding MFPT values described above. This corresponds to ~0.02 $k_B T$ difference in the free energy, which is much smaller than the calculation error, ±0.1 $k_B T$. According to Ref. [23] the atom attachment rate $f^+(n)$ can also be obtained from MFPT and $P_{larg}^{st}(n)$ (also see Supplementary Material). Figure 3(b) shows the attachment rate as a function of size $n$, with the solid line fit to $f^+(n) = D_0 n^{2/3}$ [32].

Table I gives a summary of all quantities for characterizing the nucleation process. Using the values of $\Delta G^*/k_B T$, $\rho_L$, $Z$ and $f_{n^*}^+$, we can calculate the nucleation rate $J$ according to Eq. (1) and (2), and they are very close to the values obtained directly by fitting to the MFPT curves ($J_{MFPT}$). To evaluate the accuracy of our data and the CNT predictions we performed the following calculations. First, the attachment rate can be also evaluated by measuring the effective diffusion constant at the critical nucleus size [28]. The obtained attachment rates, $f_D^*$ are close to the values obtained from the MFPT method (see Table I). Second, using $n^*$ obtained from the MFPT method, the free-energy barrier within the CNT can be calculated according to the spherical nucleus shape assumption [33]. The CNT predicts a lower barrier ($\Delta G^*_{CNT}$ in Table I) for both systems but still $\Delta G^{*Ni}_{CNT} > \Delta G^{*Ni_{50}Al_{50}}_{CNT}$. However, the CNT suggests much higher nucleation rates for both cases and most importantly $J^{Ni}_{CNT} < J^{Ni_{50}Al_{50}}_{CNT}$ which contradicts our MD simulation observation. Furthermore, the $\Delta G(n)$ profiles showed in Fig. 3 (a) is not well fitting to the traditional CNT described, i.e. ($-an + bn^{2/3}$), but could be significantly improved by adding a curvature correction term ($n^{1/3}$). Similar results also reported in other nucleation studies [32, 34, 35], and our results indicate that the nucleus shape may seriously affect the energy barrier especially when the critical nucleus size is pretty small.

The MD simulation shows how pronounced the effect of the attachment kinetics can be: the nucleation barrier is higher in the pure Ni than that in the $Ni_{50}Al_{50}$ alloy and yet the nucleation rate is 10 times higher in the pure Ni. This effect of the attachment kinetics is even more pronounced in the $Cu_{50}Zr_{50}$ alloy. Based on the data presented in Fig. 1, the nucleation barrier in the $Cu_{50}Zr_{50}$ alloy should be between those in the pure Ni and the $Ni_{50}Al_{50}$ alloy. These estimations are made based on the CNT and as we showed above can be considerably different from the actual values. However, if we assume that the CNT at least reproduces the correct trend we should observe comparable nucleation rates in these alloys. In reality, no nucleation was observed in the $Cu_{50}Zr_{50}$ alloy even at higher undercoolings ($\Delta T >$ 600 K). Therefore, the atom attachment kinetics in the $Cu_{50}Zr_{50}$ alloy should be much slower than that in the $Ni_{50}Al_{50}$ alloy. To verify this point we need to compare the atom attachment rate at the critical size in both alloys. Unfortunately, since the B2 phase never nucleates in the $Cu_{50}Zr_{50}$ alloy in the MD simulation, we cannot directly evaluate the nucleation quantities in this alloy using the methods described above. Therefore, we employed the isoconfigurational method proposed in [36] to get a rough estimation of the critical nucleus size and the effective diffusion constant $f_D^*$ at this size. In order to do it we inserted seeds of the B2 phase of different sizes in the $Cu_{50}Zr_{50}$ liquid model containing ~ 16 000 atoms at $T$ = 786K ($\Delta T$ = 540K) and performed 30 independent MD runs for each seed. The B2 phases seed with size of 32 atoms showed a 50/50 chance to melt or grow and, hence, was considered as the critical size nucleus. The effective diffusion constant [28] at this size was found to be $f_D^*$ = 4.52×10$^{10}$ s$^{-1}$. Thus, the ratio of the critical sizes obtained from the MD simulation ($n^*_{Ni} > n^*_{Cu_{50}Zr_{50}} > n^*_{Ni_{50}Al_{50}}$) does coincide with the trend predicted by the CNT and it is the extremely low diffusion constant in the $Cu_{50}Zr_{50}$ alloy (which is around 2 orders of magnitude lower than that in the $Ni_{50}Al_{50}$ alloy) which makes the nucleation is so slow.

The slower attachment kinetics in the studied alloys is probably associated with the ordered character of the B2 phase: much less fraction of the atomic jumps from the liquid phase to the growing crystal will be successful because an atom should jump into the "right" site. Of course, from time to time, an atom can jump into a "wrong" site and form an anti-site defect and MD simulation shows that the $Ni_{50}Al_{50}$ B2 phase grows from the liquid phase with a considerable amount of such defects [37]. We can speculate that the less growing crystal phase is tolerant to such defects the faster the attachment kinetics should be. To test this assumption, we compared the density of anti-site defect in $Ni_{50}Al_{50}$ and $Cu_{50}Zr_{50}$ alloys obtained by the seeding method. In both cases the seeds were larger than the critical nuclei such that the B2 phase grew from the liquid, and only the data from the as grown B2 phase were considered. Examples of the cross-section views are shown in the Fig. 4. Vividly, the growing B2 phase in the $Cu_{50}Zr_{50}$ alloy is more ordered than that in the $Ni_{50}Al_{50}$ alloy. Indeed, it was found that the new forming B2 phase in the $Ni_{50}Al_{50}$ alloy contains 15.2% anti-site defects, while their concentration in the $Cu_{50}Zr_{50}$ alloy is only 9.1%. The anti-site defect concentration in the B2 phase which spontaneously nucleated in the $Ni_{50}Al_{50}$ alloy was found to be 15.8%. The similarity of the anti-site defect concentrations obtained from the nucleation and seeded simulation shows that the atomic ordering does not come from the artificial seeds, but from the physical nature of the stoichiometric compound. Therefore, the $Cu_{50}Zr_{50}$ B2 phase is much less tolerant to the anti-site defects than the $Ni_{50}Al_{50}$ B2 phase. This strong atomic ordering results in the lower atom attachment rate, and hence, a low nucleation rate in the $Cu_{50}Zr_{50}$ alloy.

In summary, we employed MD simulation to study the crystal nucleation from the liquid in the pure Ni, $Ni_{50}Al_{50}$ and $Cu_{50}Zr_{50}$ alloys. Only the first two systems exhibited the nucleation in the course of the MD simulation. The MFPT provides a straightforward method of determining all the nucleation quantities directly from the data of the MD simulation. The analysis of these data revealed that in spite of the fact that the nucleation barrier is higher in the pure Ni the nucleation rate is also higher in the pure Ni. This was attributed to the slow atom attachment kinetics in the $Ni_{50}Al_{50}$ alloy which was related to the ordered nature of the B2 phase. The even lower fraction of the anti-site defects in the $Cu_{50}Zr_{50}$ alloy explains why the nucleation is never observed for this alloy in the course of the MD simulation. This is consistent with the experimental facts that $Cu_{50}Zr_{50}$ alloy is a good glass forming alloy and the $Ni_{50}Al_{50}$ alloy is not. The current paper demonstrates that the atom attachment rate can be the critical factor to limit the nucleation process under certain conditions and suggests a new direction in the future nucleation studies in the alloys with stoichiometric compounds focusing on their tolerance to the anti-site defects.

**Acknowledgements**

We thank M. J. Kramer, R. E. Napolitano, X. Song and R. T. Ott from Ames Laboratory for valuable discussion. Work at Ames Laboratory was supported by the US Department of Energy, Basic Energy Sciences, Materials Science and Engineering Division, under Contract No. DEAC02-07CH11358.

Table I. Summary of calculated quantities for pure Ni and $Ni_{50}Al_{50}$ ($\Delta T = 540$ K).

| Quantity | Ni | $Ni_{50}Al_{50}$ |
| --- | --- | --- |
| $N$ | 32 000 | 54 000 |
| $T$ | 1188 K | 1281 K |
| $\rho$ | 8.40e+28 $(m^3)^{-1}$ | 7.66e+28 $(m^3)^{-1}$ |
| $n^*_{MFPT}$ | 53±1 | 28±1 |
| $n^*$ | 55 | 24 |
| $\Delta G^*/k_BT$ | 19.6±0.1 | 17.9±0.1 |
| $\Delta G^*_{CNT}/k_BT$ | 12.73 | 9.20 |
| $f^+_{n^*}$ | (1.07±0.6)e+14 $s^{-1}$ | (1.26±0.6)e+12 $s^{-1}$ |
| $f^*_D$ | (9.56±0.1)e+13 $s^{-1}$ | (3.7±0.2)e+12 $s^{-1}$ |
| $Z$ | 0.0178±0.0004 | 0.032±0.001 |
| $J$ | 4.97e+32 $(m^3 \cdot s)^{-1}$ | 4.71e+31 $(m^3 \cdot s)^{-1}$ |
| $J_{MFPT}$ | (4.47±0.02)e+32 $(m^3 \cdot s)^{-1}$ | (4.69±0.02)e+31 $(m^3 \cdot s)^{-1}$ |
| $J_{CNT}$ | 5.36e+35 $(m^3 \cdot s)^{-1}$ | 1.45e+36 $(m^3 \cdot s)^{-1}$ |

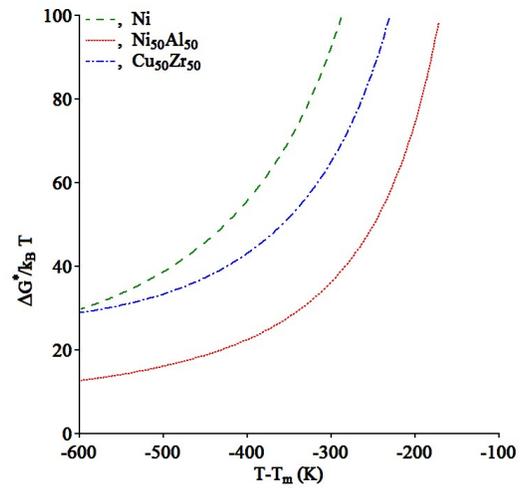

Figure 1. Nucleation barriers according to Eq. (3) as function of undercooling in the pure Ni and in the $Ni_{50}Al_{50}$ and $Cu_{50}Zr_{50}$ alloys.

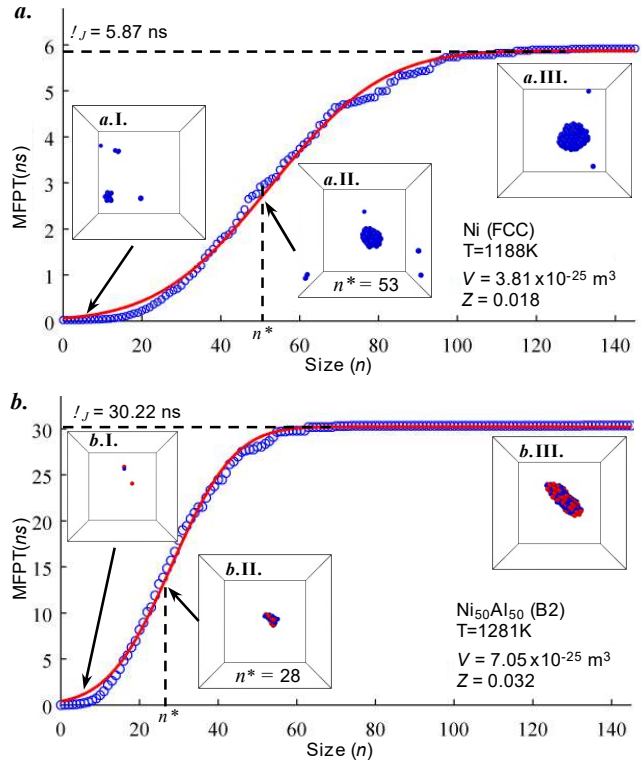

Figure 2. Mean first-passage time $\tau(n)$ as a function of the cluster size $n$ obtained from the MD simulations of (a) pure Ni and (b) Ni$_{50}$Al$_{50}$ alloys at the same undercooling $\Delta T = 540$ K. The solid lines (red) are fitted to Eq. (4). The insets show crystal clusters (I) before the nucleation, (II) when the clusters reached the corresponding critical sizes and (III) when the clusters grow ten times of their critical sizes. Ni atoms are blue and Al atoms are red. Only the solid-like atoms that determined by BOO are shown.

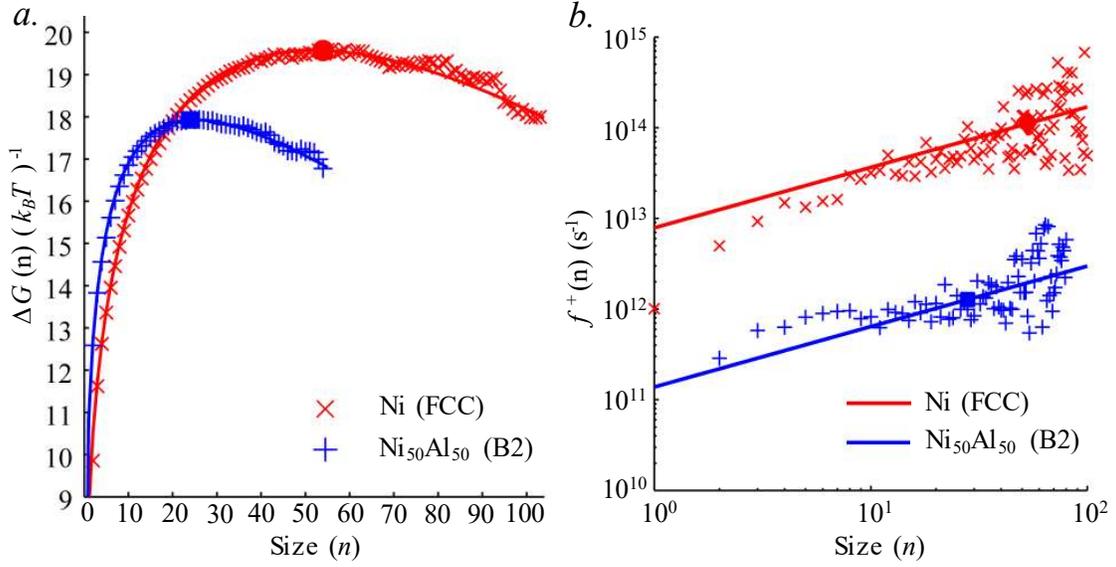

FIG. 3 (a) The true free energy curve $\Delta G(n)/k_BT$ is calculated by the Wedeking and Reguera's method and shifting according to the system size $N$. The solid points show the critical nucleation barriers. The solid lines are the fitting to $-an + bn^{2/3} + cn^{1/3}$. (b) The log-log plot of the attachment rate as a function of size $n$. The solid lines are the fitting to $f^+(n) = D_0 n^{2/3}$, and the solid points are the attachment rates at the critical size $f^+_{n^*}$.

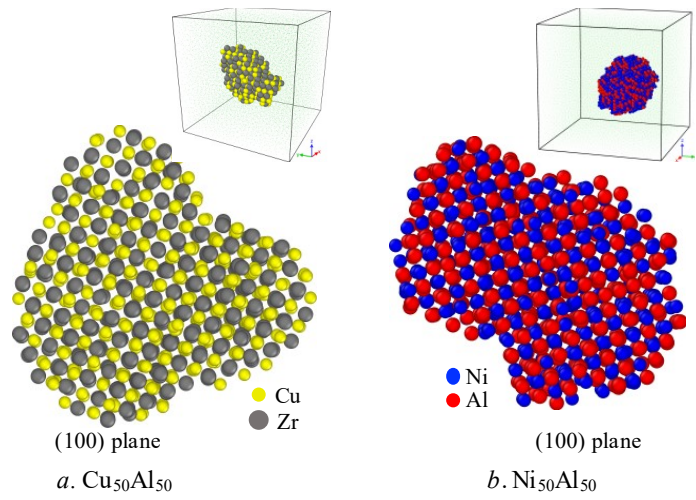

FIG. 4 The cross-section views of the (100) planes of growing B2 phases in a) the $Cu_{50}Zr_{50}$ and b) $Ni_{50}Al_{50}$ alloys.

# Supplementary Material

H. Song[a], Y. Sun[a], F, Zhang[a], C.Z. Wang[a], K.M. Ho[a,b,c] and M.I. Mendelev[a]

[a] Division of Materials Sciences and Engineering, Ames Laboratory (US Department of Energy), Ames, Iowa 50011, USA

[b] Department of Physics, Iowa State University, Ames, Iowa 50011, USA

[c] Hefei National Laboratory for Physical Sciences at the Microscale and Department of Physics, University of Science and Technology of China, Hefei, Anhui 230026, China.

In this Supplementary Material, we will summarize the main concepts of the mean first-passage time (MFPT) method used in the present study. This method relies on the Fokker-Planck equation [22, 25] which relates the probability, $P$, to find a system at point $n$ of the reaction coordinate at time $t$:

$$\frac{\partial P(n,t)}{\partial t} = \frac{\partial}{\partial n}\left[D(n)e^{-\Delta G(n)/k_B T}\frac{\partial}{\partial n}\left(P(n,t)e^{\Delta G(n)/k_B T}\right)\right] = -\frac{\partial J(n,t)}{\partial n}, \quad (S1)$$

where $J(n,t)$ is the current, $D(n)$ is a generalized diffusion coefficient, which in general depends on the state of the system, $\Delta G(n)$ is the free-energy landscape, $T$ is the temperature and $k_B$ Boltzmann's constant. In the case of nucleation, the number of atoms in a nucleus can be used as $n$.

Wedekind *et al.,* [22, 38] defined the MFPT as the average time $\tau(n)$ that the system, starting out at $n_0$, needs to reach the state $n$ for the first time; it can be presented as

$$\tau(n; n_0, a) = \int_{n_0}^{n} \frac{1}{D(y)} dy\, e^{\Delta G(y)/k_B T} \int_{a}^{y} dz\, e^{-\Delta G(z)/k_B T} \quad (S2)$$

when the boundary conditions are reflecting at $a$, and absorbing at $n = b$ [22, 39]. Therefore, we recorded the time necessary for the *largest* cluster in the system to reach or exceed a given cluster size $n$ for the first time (first-passage time) and averaged this time over several repetitions to obtain the *mean* first-passage time $\tau(n)$. The MFPT as a function of the cluster size $n$ has a sigmoidal shape [30]. If the nucleation barrier is sufficiently high, $\tau(n)$ can be approximated as [22]

$$\tau(n) = \frac{\tau_J}{2}\left[1 + erf\left(Z\sqrt{\pi}(n - n^*)\right)\right], \quad (S3)$$

where $erf(x)$ is the error function and $\tau_J$ is the nucleation time, $n^*$ is the critical cluster size and $Z$ is the Zeldovich factor. All three quantities can be obtained by fitting the MFPT data to Eq. (S3).

The nucleation rate can be obtained from the nucleation time $\tau_J$ as

$$J = \frac{1}{V\tau_J}. \tag{S4}$$

To determine the free-energy barrier for the nucleation, we used the method developed by Wedekind and Reguera [23, 30]. This method required only two ingredients: the steady-state probability distribution and the MFPT to reconstruct the free-energy landscape. Both ingredients can be directly obtained from the MD simulation. In particular, Wedekind *et. al.* choose the largest embryo in the system as the appropriate order parameter to track the MFPT $\tau(n)$ and steady-state probability distribution $P_{larg}^{st}(n)$. Following [23, 30], we first calculated:

$$B(n) = -\frac{1}{P_{larg}^{st}(n)} \left[ \int_n^b P_{larg}^{st}(n')dn' - \frac{\tau(b)-\tau(n)}{\tau(b)} \right], \tag{S5}$$

where $b$ is an upper boundary that we sample both for $\tau(n)$ and $P_{larg}^{st}(n)$, which means that once the largest embryo has passed through this boundary in a simulation the following time steps are discarded from the statistics. We chose $b = 140$ for the pure Ni, and $b = 70$ for $Ni_{50}Al_{50}$. Both of the $b$ boundaries are more than twice of the critical nucleus sizes in each system. Then, the free energy $\Delta G_{larg}(n)$ is reconstructed by

$$\Delta G_{larg}(n)/k_B T = \ln\left(\frac{B(n)}{B(n_1)}\right) - \int_{n_1}^{n} \frac{dn'}{B(n')} + C, \tag{S6}$$

for any desired interval $[n_1 \leq n \leq b]$. As a reference point we used $n_1 = 1$, and the constant C is applied to match up curve $\Delta G_{larg}(n)/k_B T$ with $-\ln(P_{larg}^{st}(n))$ in certain $n$ range (See below).

The equilibrium distribution of embryos can be described by [2, 40]:

$$P^{st}(n) = \frac{N_n}{N} = exp\left[-\frac{\Delta G(n)}{k_B T}\right], \tag{S7}$$

where $N_n$ is the number of embryos of size $n$ in a system containing $N$ atoms. The steady-state probability distribution $P^{st}(n)$ for any embryo size $n$ can be approximated by $P^{st}(n) = N_n/N$, However, the probability distribution of the largest embryo $P_{larg}^{st}(n)$ is *unequal* to the full probability distribution of all the embryos $P^{st}(n)$ in the system [31]. Instead, $P_{larg}^{st}(n)$ should be considered as the probability of forming exactly *one* (the largest) embryo in the system. Thus, both $P_{larg}^{st}(n)$ and $P^{st}(n)$ are related in a similar fashion by $\frac{P_{larg}^{st}(n)}{N} \approx P^{st}(n)$, and the free energies for the largest or any cluster are connected by the simple relation $\Delta G(n) = \Delta G_{larg}(n) + k_B T\ln(N)$ [30]. Lundrigan and Saika-Voivod indicated that in a high free-energy barrier nucleation, larger embryos are rare, when a large embryo is present, there is approximately no other embryo of that size or larger in the system [32]. Therefore, for large embryo size $\frac{P^{st}_{larg}(n)}{N}$ and $P^{st}(n)$ are approximately equal to each other. The curves $\Delta G_{larg}(n)/k_B T$ and $-\ln(P^{st}_{larg}(n))$ are shown in Figs. S1a and S1c for the pure Ni and Ni$_{50}$Al$_{50}$ alloy, respectively. According to Eq. (S7), $-\ln(P^{st}_{larg}(n))$ can be a reference for the nucleation free-energy for small cluster sizes, since the small clusters are nearly equilibrium. However, upon approaching the critical size and beyond it, the continuously upward the $-\ln(P^{st}_{larg}(n))$ curve shows that the free-energy landscape cannot be recovered by simply using Eq. S7 since the formation of a cluster in MD is not an equilibrium process and the probability distribution we collected from the simulation is not the equilibrium one. Consequently, we used only the $-\ln(P^{st}_{larg}(n))$ curve at small $n$ as the reference to line up the $\Delta G_{larg}(n)/k_B T$ curve by the constant $C$ in Eq. S6. Figures S1a and S1c show an excellent agreement between the $\Delta G_{larg}(n)/k_B T$ and $-\ln(P^{st}_{larg}(n))$ curves at small $n$ after aligning by $C$. Now the true free-energy landscapes $\Delta G(n)/k_B T$ can be obtained by shifting the $\Delta G_{larg}(n)/k_B T$ curves by $\ln(N)$ which is shown in Figs. S1b and S1d. Lundrigan and Saika-Voivod used the Monte Carlo (MC) simulation with the umbrella sampling method [20, 28, 41, 42] to verify the free-energy landscapes obtained by the MFPT method. They pointed out that although there is a minor deviation in the small cluster size range, both methods show a very good agreement at the nucleation energy barrier [32].

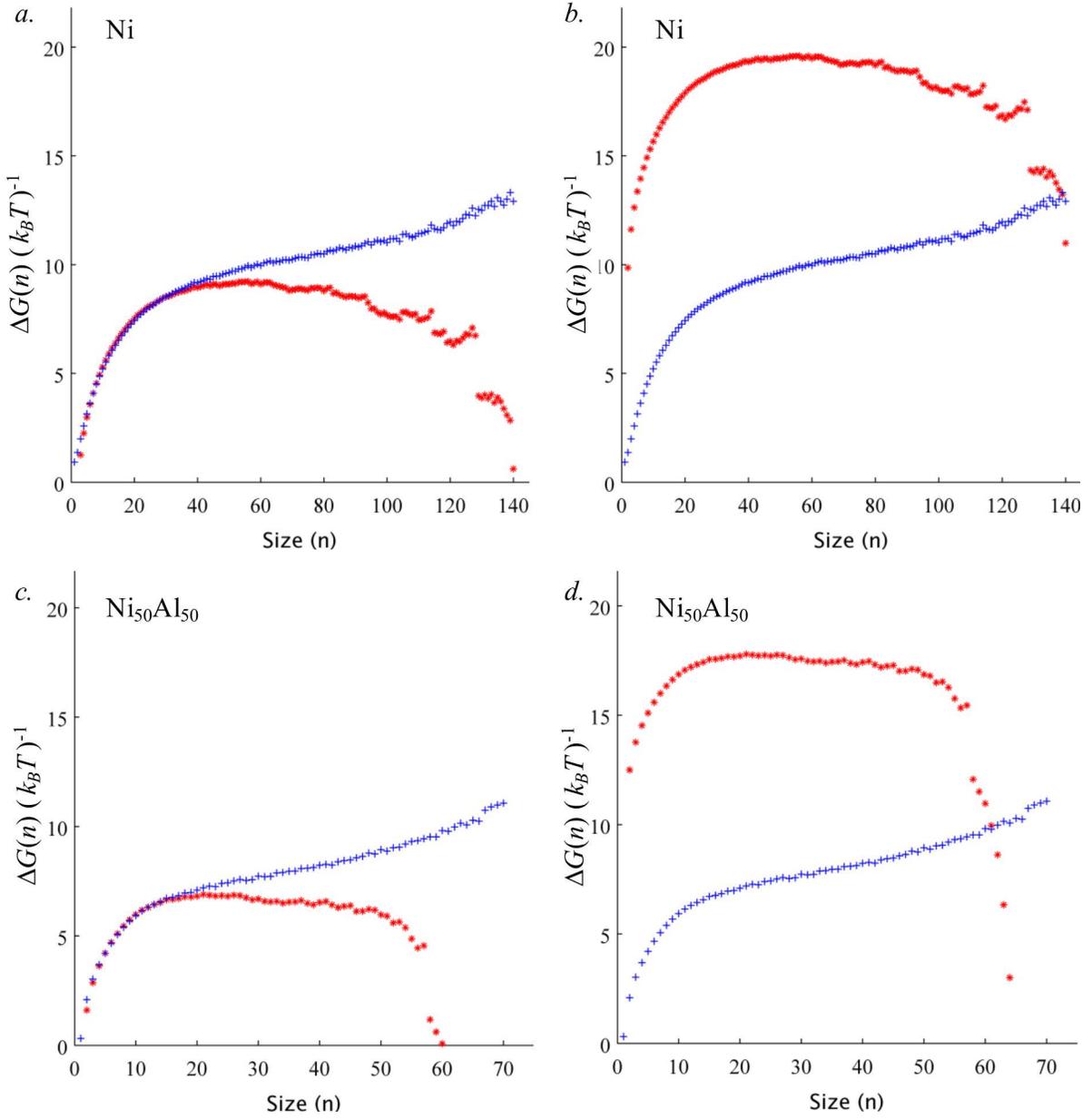

*Figure S1. Free-energy landscape for the nucleation in the a,b) pure Ni and c,d) $Ni_{50}Al_{50}$ alloy. Red stars represent $\Delta G_{larg}(n)/k_BT$, and blue crosses represent $-\ln(P^{st}_{larg}(n))$. In a) and c), $\Delta G_{larg}(n)/k_BT$ curves are lined up with $-\ln(P^{st}_{larg}(n))$ in the region before the critical sizes by the constant C in Eq. S6. Plots in b) and d) show the true free-energy landscapes by shifting the $\Delta G_{larg}/k_BT$ curves by $\ln(N)$ in each system.*

The atom attachment rates as a function of *n* can be obtained from the MFPT and $P^{st}_{larg}(n)$ data [22, 30, 32] as:

$$f^+(n) = B(n)/\left(\frac{\partial \tau(n)}{\partial n}\right). \tag{S8}$$

We assume that the rate of attachment of an atom to the critical cluster is the same for both any and the largest clusters. We also applied the method developed by Auer and Frenkel [28] to calculate the single point for $f_D^*$ at the critical nucleus size $n^*$. Figures S2a and S2b (also see Table I in the main text) show that although there is some noise from $P_{larg}^{st}(n)$ at the large cluster size, the agreement near $n^*$ between the two methods is acceptable.

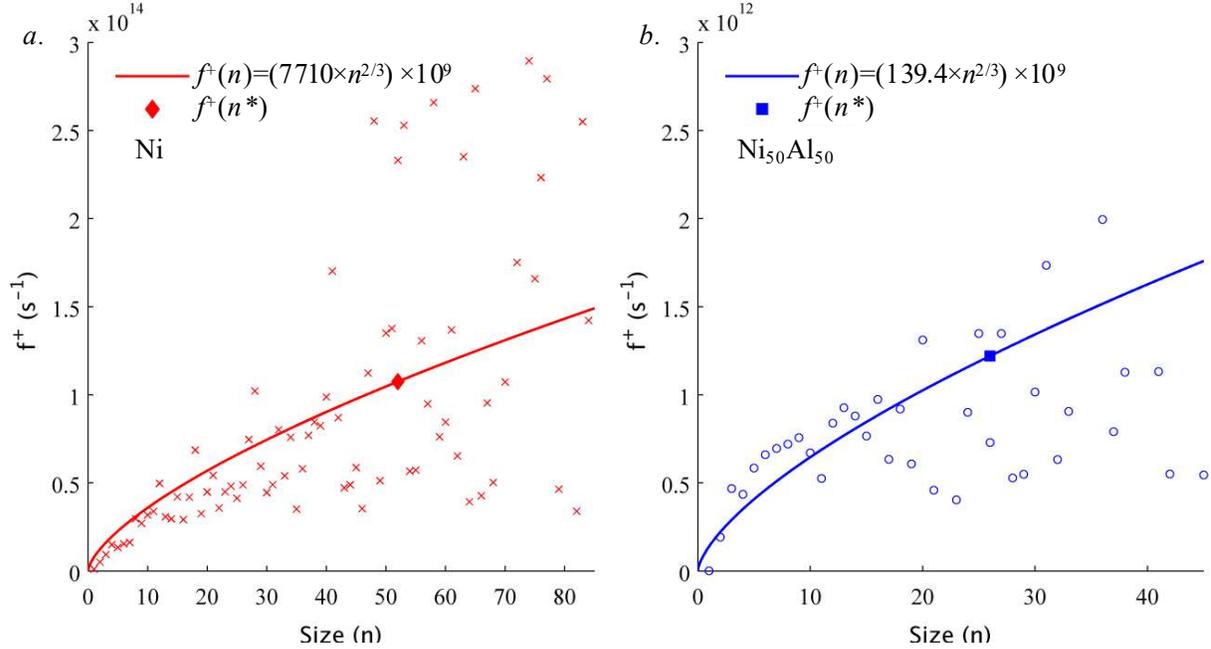

Figure S2. The atom attachment rates as a function of the cluster size n in the a) pure Ni and b) $Ni_{50}Al_{50}$ alloy. The solid lines are the fittings to $f^+(n) = D_0 \cdot n^{2/3}$, and the solid points show the attachment rates $f^+(n^*)$ at the critical nucleus size $n^*$.